.% ****** Start of file apssamp.tex ******
\documentclass[aip,reprint,onecolumn,showpacs,preprintnumbers,amsmath,amssymb]{revtex4-1}
\usepackage[dvips]{graphicx}
\usepackage{psfrag}
\usepackage{epsfig}
\usepackage{dcolumn}% Align table columns on decimal point
\usepackage{bm}% bold math
\usepackage{color}
\usepackage{paralist}
\usepackage{hyperref}
\usepackage[usenames,dvipsnames]{xcolor}
\newcommand{\avg}[1]{\overline{#1}}
\newcommand{\spavg}[1]{\left\langle{#1}\right\rangle}
\newcommand{\vex}[1]{\mathbf{#1}}
\newcommand{\vu}{\vex{u}}

\newcommand{\ten}[1]{\mathbf{#1}}
\newcommand{\tT}{\ten{T}}
\newcommand{\tS}{\ten{S}}
\newcommand{\tC}{\ten{C}}

\newcommand{\vnabla}{\boldsymbol{\nabla}}
\newcommand{\pdert}[1]{\partial_t #1}

\newcommand{\der}[2]{\frac{d #1}{d #2}}
\newcommand{\eqnref}[1]{Eq.~(\ref{#1})}
\newcommand{\figref}[1]{Figure~\ref{#1}}
\newcommand{\figrefalt}[1]{Figure~\ref{#1}}

\newcommand{\etal}{\textit{et al.}}
\newcommand{\ie}{\textit{i.e.}}
\def\square{${\vcenter{\hrule height .8pt
        \hbox{\vrule width .8pt height 5pt \kern 5pt
        \vrule width .8pt}
        \hrule height .8pt}}$}

\newcount\ndots
\def\drawline#1#2{\raise 2.5pt\vbox{\hrule width #1pt height #2pt}}
\def\spacce#1{\hskip #1pt}
\def\solid{\drawline{24}{.5}\nobreak\ }
\def\boldsolid{\drawline{24}{2.}\nobreak\ }
\def\bdash{\hbox{\drawline{4}{.5}\spacce{2}}}
\def\dashed{\bdash\bdash\bdash\bdash\nobreak\ }

\def\bdot{\hbox{\drawline{1}{.5}\spacce{2}}}
\def\dotted{\hbox{\leaders\bdot\hskip 24pt}\nobreak\ }

\def\chndot{\hbox {\drawline{9.5}{.5}\spacce{2}\drawline{1}{.5}\spacce{2}\drawline{9.5}{.5}}\nobreak\ }

\def\trian{\raise 1.25pt\hbox{$\scriptscriptstyle\triangle$}\nobreak\ }
\def\circle{$\circ$\nobreak\ }

\def\solidcircle{$\bullet$\nobreak\ }

\def\square{${\vcenter{\hrule height .4pt
        \hbox{\vrule width .4pt height 3pt \kern 3pt
        \vrule width .4pt}
        \hrule height .4pt}}$\nobreak\ }
\def\plus{\raise 1.25pt \hbox{$\scriptscriptstyle +$}\nobreak\ }
\def\x{\raise 1.25pt \hbox{$\scriptscriptstyle \times$}\nobreak\ }

\def\solidtrian{\raise 1.25pt
   \hbox to 3bp{\special{ newpath  0 0 moveto 3 0   lineto 1.5 2.598
lineto closepath fill}\hfill}\nobreak\ }
\def\solidsquare{\vrule height .9ex width .8ex depth -.1ex\nobreak\ }

\def\solidcclose{\drawline{10}{.5}\nobreak\raise
  0.5pt\hbox{$\bullet$}\drawline{10}{.5}\nobreak\ }
\def\solidsclose{\drawline{10}{.5}\nobreak\raise
  0.5pt\hbox{\solidsquare}\drawline{10}{.5}\nobreak\ }
\def\solidtclose{\drawline{10}{.5}\nobreak\raise
  0.5pt\hbox{\solidtrian}\drawline{10}{.5}\nobreak\ }
\def\solidcopen{\drawline{10}{.5}\nobreak\raise
  0.5pt\hbox{\circle}\drawline{10}{.5}\nobreak\ }
\def\solidsopen{\drawline{10}{.5}\nobreak\raise
  0.5pt\hbox{\square}\drawline{10}{.5}\nobreak\ }
\def\solidtopen{\drawline{10}{.5}\nobreak\raise
  0.5pt\hbox{\trian}\drawline{10}{.5}\nobreak\ }
\def\solidx{\drawline{10}{.5}\nobreak\raise
  0.5pt\hbox{\x}\drawline{10}{.5}\nobreak\ }
\begin{document}

\preprint{APS/123-QED}

\title{On the mechanism of elasto-inertial turbulence
}% Force line breaks with \\

\author{Yves Dubief}
 \email{yves.dubief@uvm.edu}
% \homepage{http://www.uvm.edu/~ydubief}
\affiliation{
School of Engineering,
 University of Vermont,
 Burlington VT}
 \author{Vincent E. Terrapon}
\affiliation{Aerospace and Mechanical Engineering Department, University of Liege, Belgium}
\author{Julio Soria}
\affiliation{Department of Mechanical and Aerospace Engineering, Monash University, Australia, and
Department of Aeronautical Engineering, King Abdulaziz University, Jeddah, Kingdom of Saudi Arabia}%

\date{\today}% It is always \today, today,
\begin{abstract}
Elasto-inertial turbulence is a new state of turbulence found in flows with polymer additives .
The dynamics of turbulence generated and controlled by such additives is investigated from the perspective of the coupling between polymer dynamics and flow structures. Direct numerical simulations of channel flow with Reynolds numbers ranging from 1000 to 6000 (based on the bulk and the channel height) are used to study the formation and  dynamics of elastic instabilities and their effects on the flow. The resulting mechanism of interactions between polymer dynamics and the flow helps resolve a long-standing controversy in the understanding of polymer drag reduction and explains the phenomenon of early turbulence, or onset of turbulence at lower Reynolds numbers than for Newtonian flows, previously observed in polymeric flows. Polymers also point out an interesting analogy with the forward and backward energy cascade in two-dimensional turbulence.

\end{abstract}

\maketitle

\section*{Foreword}
This article is dedicated to our mentor, colleague and friend, Prof. Parviz Moin, in honour of his 60th birthday, and without whom this research would have never been undertaken. First, because our interest in polymeric flows started through Parviz' impulsion and guidance more than ten years ago, when research on turbulent drag reduction by polymer additives was initiated at the Center for Turbulence Research (CTR) in Stanford, and which project we had the privilege to be part of. Secondly, because years later the umbrella of the CTR and its Summer Program gave us the chance to revisit and further deepen this topic. The collaborative and incredibly fertile atmosphere of the CTR, which Parviz created and nurtured, was indeed a major contributor to the results presented here. Finally, and foremost, because Parviz' vision and pioneering work in numerical simulations of turbulence paved the way for so many future research in this field. This is very well exemplified by the results presented here, which demonstrate not only the capacity of high-fidelity computations to accurately simulate experimental observations, but also to offer an unparalleled insight into the physics of such complex flows. We are forever in debt to Parviz for his tremendous help, support and inspiration.

%=========================================================================
\section{Introduction}
%=========================================================================
Polymer additives are known for producing upward of 80\% of drag reduction in turbulent wall-bounded flows through a strong alteration and reduction of the turbulent activity \cite{white2008map}. The changes in flow dynamics induced by polymers do not lead to flow relaminarization but, at most, to a universal asymptotic state called maximum drag reduction (MDR). The early quantitative description of MDR \cite{virk1970uaa} was semi-empirical, yielding a correlation for the friction factor and the Virk log-law mean velocity profile.  Recently,  Procaccia \etal\cite{procaccia2008colloquium} proposed a theory that derives, in the limit of infinite Reynolds number, an asymptotic log-law for the mean velocity profile remarkably close to Virk's. However, at low and moderate Reynolds numbers,the existence of a logarithmic region in the mean velocity profile of MDR is not verified by currently available experiments and simulations \cite{white2012re}.

The theory of MDR derived by Procaccia \textit{et al.} describes the action of polymers as an eddy viscosity linearly increasing with increasing distance to the wall. Most noticeably, the theory assumes that ``In contradistinction to the picture offered by de Gennes, simulations (performed by the authors) indicate that the energy never goes back from the polymers to the flow; the only thing that polymers can do is to increase the dissipation'' (From Procaccia \etal\cite{procaccia2008colloquium}). This statement is at odds with the mechanism of polymer drag reduction and two other unique properties of polymer solutions,  elastic turbulence and elasto-inertial turbulence. These three phenomena support de Gennes\cite{de1990introduction}'s picture that drag reduction derives from two-way energy transfers between turbulent kinetic energy of the flow and elastic energy of polymers at small scales, resulting into the overall modification of the turbulence energy cascade. 

The drag reducing mechanism is caused by an increase of the (extensional) viscosity in extensional upwash and downwash flows generated by quasi-streamwise vortices \cite{dubief2004cdr,terrapon2004sps}, thereby creating a negative torque on these near-wall vortices \cite{kim2007effects}. Dubief \etal\cite{dubief2004cdr} demonstrated that polymers re-inject part of the energy accumulated  in high-speed streaks, regions of locally high-speed flow in the near-wall region elongated in the direction of the flow. In inertia-less flows with curved streamlines, Groisman \& Steinberg\cite{groisman2000elastic} demonstrated the existence of strong nonlinear mixing supported by elastic turbulence, a state of saturated dynamical interactions between stretched polymer molecules and the base flow that causes the stretching. 

The present paper focuses on the latest evidence of energy transfer from polymers to flow: the recently discovered elasto-inertial turbulence, hereafter referred to as EIT \cite{samanta2012transition}. EIT is a new state of small-scale turbulence driven by the interaction between elastic instabilities and the flow's inertia that has been observed over a wide range of Reynolds numbers, from subcritical to supercritical Reynolds numbers. EIT exists by either creating its own extensional flow patterns, as we will demonstrate here in subcritical channel flows, or by exploiting extensional flow topologies, as we shall observe in the wake of a hairpin vortex at higher Reynolds numbers. EIT offers an alternative description of MDR that has the merit of explaining  the phenomenon of early turbulence \cite{hoyt1977laminar}, which describes the onset of turbulence in the presence of diluted polymer additives at Reynolds numbers significantly smaller than in the absence of polymers. EIT also provides a unique insight on the possible coexistence of inverse energy cascade in three-dimensional flows.

\section{Method}
Channel flow simulations are performed in a cartesian domain, where $x$, $y$
and $z$ are the streamwise, wall-normal and spanwise directions,
respectively. For a polymer solution, the flow transport
equations are the  conservation of mass, $\vnabla\cdot\vex{u}=0$, where $\vu$ is the velocity vector,
and transport of momentum:
\begin{equation}
\pdert{\vex{u}}+(\vex{u}\cdot\vnabla)\vex{u}=-\vnabla p+\frac{\beta}{Re}\nabla^2\vex{u}+\frac{1-\beta}{Re}\vnabla\cdot\tT\,. \label{eq:mom}
\end{equation}
The Reynolds number is based on the bulk velocity $U_b$ and the full channel height $H=2h$, $Re=U_bH/\nu$.
The parameter $\beta$ is the ratio of solvent viscosity to the zero-shear viscosity of the polymer solution and affects both the viscous stress and polymer stress terms in \eqnref{eq:mom}. The polymer stress tensor $\ten{T}$ is computed using the FENE-P (Finite Elastic Nonlinear Extensibility-Peterlin) model \citep{bird1987dynamics}:
\begin{equation}
\ten{T}=\frac{1}{Wi}\left(\frac{\ten{C}}{1-\text{tr}(\ten{C})/L^2}-\ten{I}\right)\label{eq:taup}\;,
\end{equation}
where the tensor $\ten{C}$ is the local conformation tensor of the polymer solution and $\ten{I}$ is the unit tensor. The properties of the polymer solution are $\beta$, the maximum polymer extension $L$, and the Weissenberg number $Wi$ based on the solution relaxation time $\lambda$ and the flow time scale relevant to the dynamics of interest. Here $Wi$ is based on the wall shear-rate $\dot{\gamma}$ of the initial laminar flow at each $Re$, hence $Wi=\lambda\dot{\gamma}$.  The FENE-P model assumes that polymers may be represented by a pair of beads connected by a nonlinear spring defined by the end-to-end vector $\vex{q}$. The conformation tensor is the phase-average of the tensorial product of the end-to-end vector $\vex{q}$ with itself, $\ten{C}=\langle\vex{q}\otimes\vex{q}\rangle$, whose transport equation is
\begin{equation}
\pdert{\ten{C}}+(\vex{u}\cdot\vnabla)\ten{C}=\ten{C}(\vnabla\,\vex{u})+(\vnabla\,\vex{u})^\text{T}\ten{C}-\tT\;. \label{eq:C}
\end{equation}
On the right-hand side of \eqnref{eq:C}, the first two terms are
responsible for the stretching of polymers by hydrodynamic forces,
whereas the third term models the internal energy that tends to
bring stretched polymers to their least energetic state (coiled).

Eqs.~(\ref{eq:mom}-\ref{eq:C}) are solved using finite differences
on a staggered grid and a semi-implicit time advancement scheme described elsewhere \citep{dubief2005nai}. A series of simulations was carried out for Reynolds numbers ranging from 1000 to 6000. A thorough resolution study  led us to choose a domain size of $10H\times H\times 5H$ with $256\times151\times256$ computational nodes. All results discussed here have been verified on domains with a factor 2 in horizontal dimensions and resolution in each directions. The CFL number was set to 0.15  to guarantee the boundedness of $\ten{C}$. 

 The protocol for our simulations was designed to mimic the perturbed experimental setup of Samanta \etal\cite{samanta2012transition} within the limitation inherent to the DNS boundary conditions. For any flow, Newtonian or polymeric, the initial flow and polymer fields are first equilibrated to the laminar state corresponding to the desired Reynolds number.  A perturbation is then introduced over a short duration, in the form of blowing and suction velocity on both walls, over which white noise of prescribed intensity is introduced. The velocity pattern is periodic in $x$ and $z$:
 \begin{equation}
 v_w(x,z,t)={\cal H}(t)\left[A\sin\left(\frac{8\pi}{L_x} x\right)\sin\left(\frac{8\pi}{L_z}z\right)+\varepsilon(t)\right],\label{eq:vw}
 \end{equation}
 where A is the amplitude, $L_x$ and $L_z$ are the horizontal domain dimensions, and $\varepsilon(t)$ is the random noise. The total duration of the perturbation is $0.5h/U_b$, of which the first and last 10\% correspond to a gradual increase/decrease through a smooth step function ${\cal H}(t)$. Choosing $A=0.09U_b$ and the RMS of $\varepsilon$ at $0.005U_b$ causes the Newtonian flow to transition at $Re=6000$.

All polymeric simulations are performed with $L=200$ and $\beta=0.9$. Two Weissenberg numbers are considered, $Wi=100$ and 700. The former is consistent with previous simulations of MDR \citep{dubief2004cdr,li2006influence,white2012re}.  The theory of Procaccia \etal\cite{procaccia2008colloquium}, based on infinite $Re$ and $Wi$, motivates the second, as an exploration of the effects of very large elasticity, or $Wi$, on the flow. 

\section{Statistical description of the polymeric flows}
\subsection{Velocity statistics}
\begin{figure}
\centerline{
\includegraphics[width=0.98\textwidth]{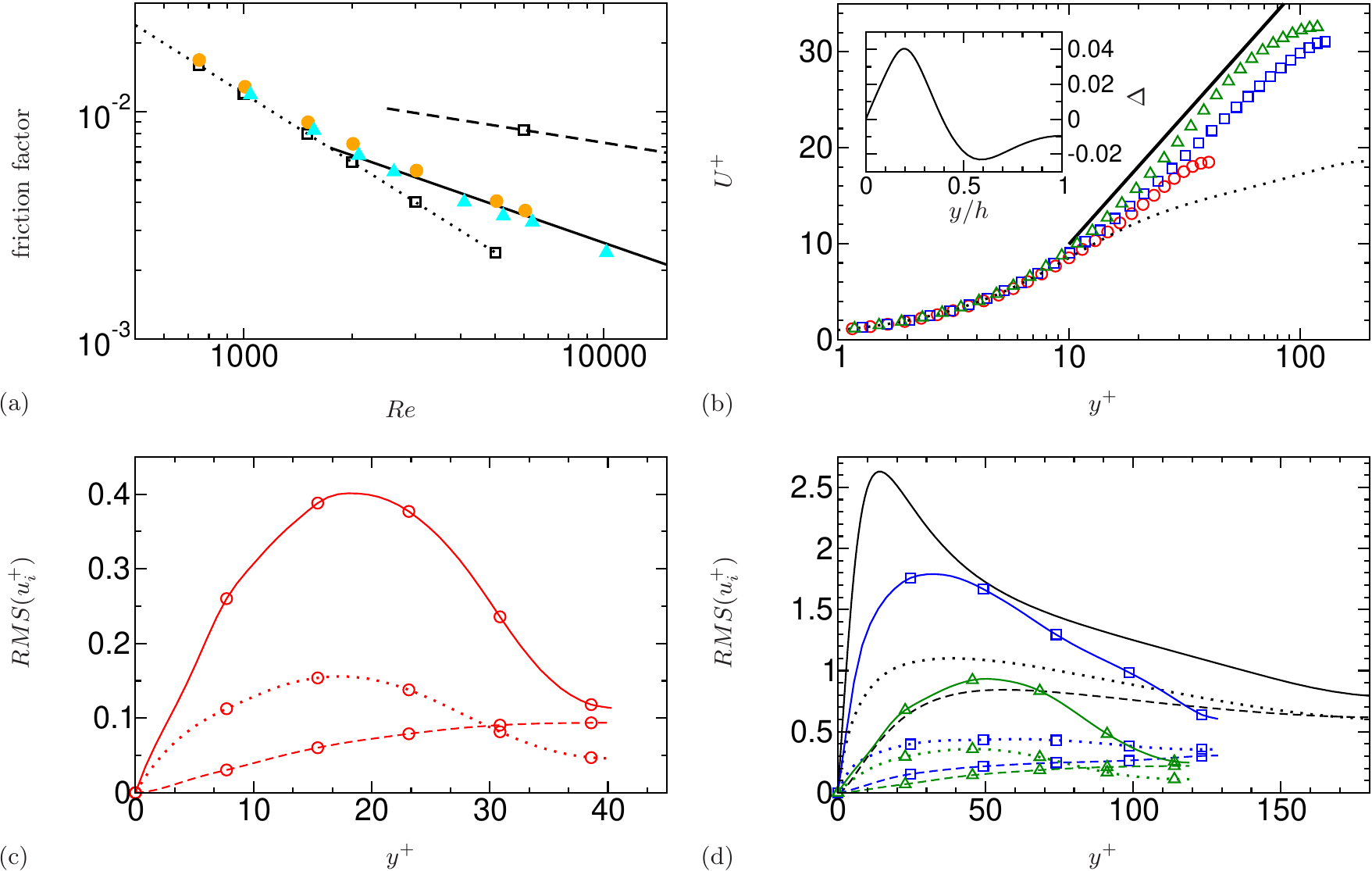}
}
\caption{\label{fig:stat} (a) Friction factor as a function of the Reynolds number for two Weissenberg numbers, $Wi=100$ ({\color{orange} $\bullet$}) and $Wi=700$  ({\color{cyan}$\blacktriangle$}), with $L=200$ and $\beta=0.9$. Lines indicate correlations for laminar (\dotted, $f=12/Re$) and turbulent (\dashed, $f=0.073Re^{-1/4}$) Newtonian channel flow and for MDR (\solid, $f=0.42Re^{-0.55}$); Newtonian solutions are also included (\square). (b) Mean velocity profiles normalized by viscous scales; polymer flows: {\color{red} $\circ$},  $Re=1000$, $Wi=100$; {\color{blue} \square}, $Re=6000$, $Wi=100$; {\color{ForestGreen} \trian}, $Re=6000$, $Wi=700$; \dotted, Newtonian flow at $Re=6000$. \boldsolid, Virk's MDR velocity profile $U^+=11.7\ln(y^+)-17$. Inset: Difference between the analytical Newtonian laminar velocity profile and the velocity profile of polymer flow at $Re=1000$, $Wi=100$.  (c) RMS of velocity fluctuations normalized by viscous scales: \solid, streamwise velocity $u'$; \dashed, wall-normal velocity $v'$; \dotted, spanwise velocity $w'$ for polymer flow at $Re=1000$, $Wi=100$.  (d) Same as (c) for polymer flows at $Re=6000$, $Wi=100$ and $Wi=700$, and for Newtonian flow at $Re=6000$ (lines without symbols). For colors and symbols see (b). 
}
\end{figure}

%%BW\begin{figure}
%%BW\vspace*{12pt}
%%BW\centerline{
%%BW\includegraphics[width=0.98\textwidth]{Fig1-bw.pdf}
%%BW}
%%BW\caption{\label{fig:stat} (a) Shows the friction factor as a function of Reynolds number for two Weissenberg numbers $Wi=100$ ({$\bullet$}), $Wi=700$  ({$\blacktriangle$}) with $L=200$ and $\beta=0.9$. Lines indicate correlations for laminar (\dotted, $f=12/Re$) and turbulent (\dashed, $f=0.073Re^{-1/4}$) Newtonian channel flow and for MDR (\solid, $f=0.42Re^{-0.55}$); Newtonian solutions are also included (\square). (b) and (c) Display the mean velocity profiles and profiles of turbulent kinetic energy, respectively, for polymeric simulations with $Wi=100$ at $Re=1000$ ({\solid}, compared to the Poiseuille laminar solution \dotted) and $Re=6000$ ({\dashed}), and for fully turbulent Newtonian flow at $Re=6000$ ({\chndot}). (d) Shows the profile of polymer extension corresponding to the polymeric flows used in (b) and (c). (Color online.)
%%BW}
%%BW\end{figure}

\begin{figure}
\centerline{
\includegraphics[width=0.98\textwidth]{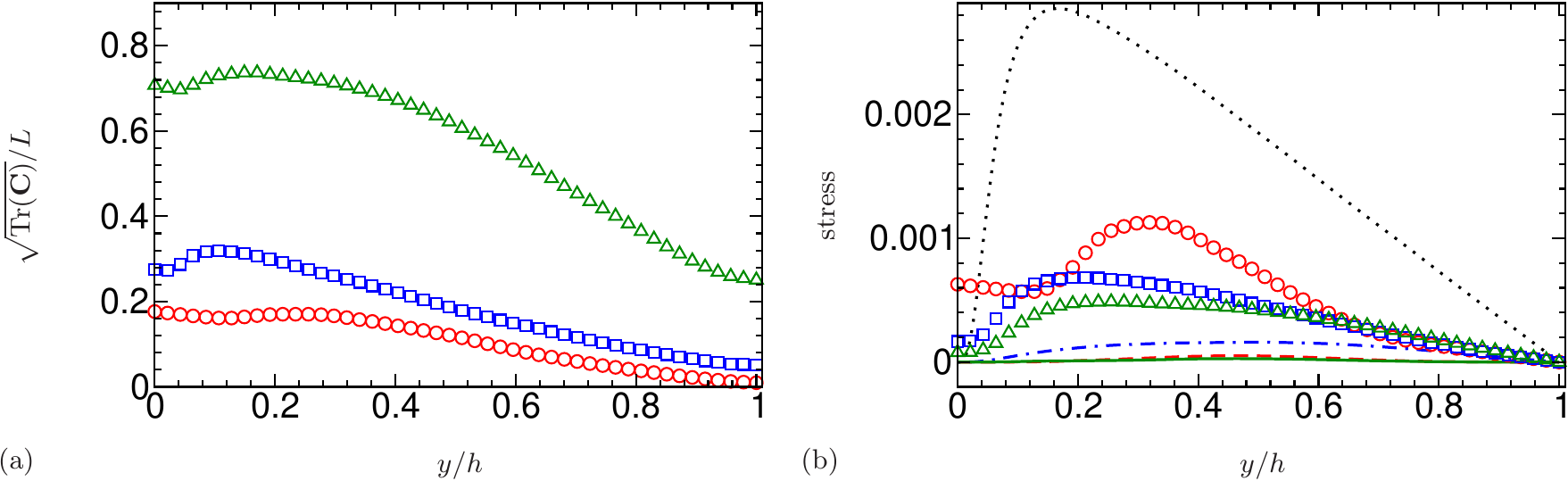}
}
\caption{\label{fig:statpoly} (a) Mean polymer extension: {\color{red} $\circ$},  $Re=1000$, $Wi=100$; {\color{blue} \square}, $Re=6000$, $Wi=100$; {\color{ForestGreen} \trian} $Re=6000$, $Wi=700$. (b) Polymer (same symbols as in a)) and Reynolds (lines) shear stresses normalized by the bulk velocity and channel half-height: {\color{red} \dashed},  $Re=1000$, $Wi=100$; {\color{blue} \chndot}, $Re=6000$, $Wi=100$; {\color{ForestGreen} \solid}, $Re=6000$, $Wi=700$; \dotted, $Re=6000$, no polymers. 
}
\end{figure}

\figrefalt{fig:stat} summarizes the main statistical properties of our simulations, which were found to be in excellent agreement with experiments in a pipe flow \citep{samanta2012transition}. \figrefalt{fig:stat}(a) shows the evolution of the friction factor as a function of the Reynolds number from $Re=1000$ to 6000. The friction factor of polymeric flows departs from the laminar asymptote, with a slight drag increase, to transition smoothly to the MDR asymptote, where drag is reduced compared to turbulent flow. The effects of increasing elasticity is only noticeable beyond the junction between the laminar and MDR asymptotes of the friction factor at $Re_c=1791$. The lower Weissenberg number simulations reach a high drag reduced state (HDR\cite{dubief2004cdr}), while the high Weissenberg number simulations achieve MDR. 

Figures \ref{fig:stat}(b-d) compare profiles of velocity statistics of the polymeric and Newtonian flows at  two Reynolds numbers, $Re=1000$ and 6000, representative of the two domains $Re<Re_c$ and $Re>Re_c$. At low Reynolds numbers, there is no evidence of logarithmic behavior in the mean velocity profile at HDR or MDR, as discussed in a previous publication\cite{white2012re}. The mean velocity profile departs from the Newtonian Poiseuille solution for laminar flow (\figrefalt{fig:stat}b) by only a few percents. In the second domain ($Re>Re_c$), mean velocity profiles differ significantly  from Newtonian turbulence at the same $Re$. The highest Weissenberg simulation approaches Virk's MDR velocity profile\cite{virk1970uaa}. When normalized by viscous flow scales \figrefalt{fig:stat}(c-d), the RMS of velocity fluctuations show a consistent picture at the two Reynolds and Weissenberg numbers considered here, that qualitatively resembles that of Newtonian wall-turbulence. The existence of a maximum in the RMS of streamwise velocity fluctuations suggest the presence of streaks in all flows. The intensity of these streaks and their thickness is obviously modulated by the elasticity of the solution, as obvious at $Re=6000$ in \figrefalt{fig:stat}(d) from the shift of the maximum of $RMS(u^+)$  from $y^+\approx12$ (Newtonian) to $y^+\approx30$ ($Wi=100$) and $y^+\approx50$ ($Wi=700$). The spatial distributions of spanwise velocity fluctuations for $Re=1000$, $Wi=100$ and $Re=6000$ and $Wi=700$ are qualitatively similar with a more significant decrease of the centerline $RMS$ relative to its maximum than observed for the Newtonian  and $Re=6000$, $Wi=700$ flows. In all polymer flows the distribution of the wall-normal velocity fluctuations is maximum at the centerline, which is a difference from Newtonian turbulence.

The similarity between $Re=1000$, $Wi=100$ and  $Re=6000$, $Wi=700$ suggests that the two flows create similar turbulent structures. We will later confirm that $Re=6000$, $Wi=700$ is indeed dominated by elasto-inertial turbulence. The difference, especially in $RMS(w^+)$, with $Re=6000$, $Wi=100$ indicates the existence of a flow which is still under some influence of typical structures found in Newtonian wall turbulence.

\subsection{Polymer statistics}
\figref{fig:statpoly}(a) shows the profiles of polymer stretch, defined as $\left(\text{tr}(\avg{C})\right)^{1/2}/L$, for the three simulations of interest. The mean polymer stretch is well below 50\% of full extension at $Wi=100$, indicating that the mechanism at play is not induced by coil-stretch transition, at least from the perspective of time-averaged transport equations. The most elastic solution has significantly larger mean stretch, yet its mean is well below 100\%. In a purely laminar flow, \ie, a pure shear flow, the solution for the polymer stretch decreases monotonically and linearly due to the linear behavior of the shear. The emergence of a local maximum away from the wall at our lowest Reynolds number suggests the existence of extensional flow topologies, as such topologies are known to produce the largest polymer extensions in wall turbulence. Indeed, the local maximum of polymer at $Re=6000$ arises from the interactions between polymers and turbulent structures in the buffer region, in particular vortices, which produce local extensional flows with dramatic effects on the polymer dynamics \cite{terrapon2004sps}. The possible existence of extensional flow topology at $Re=1000$ is the motivation for our next section.

Before studying the emergence of extensional flows, the Reynolds shear stress $-\avg{uv}$ and polymer stress $(1-\beta)\avg{T}_{xy}/Re$ terms are compared (\figref{fig:statpoly}b). We remind the reader that the two stresses, with the addition of the viscous stress (not shown), leads for the balance of stresses in a channel flow to:
\begin{equation}
-\frac{dP}{dx}\left(1-\frac{y}{h}\right)=-\avg{uv}+\frac{1-\beta}{Re}\avg{T}_{xy}+\frac{\beta}{Re}\der{U}{y}\,.
\end{equation}
For all simulations, the Reynolds shear stress is very small compared to the polymer shear stress. Supporting the proposition that flows at $Re=1000$, $Wi=100$ and  $Re=6000$, $Wi=700$ are similar, the Reynolds shear stress is negligible in both cases, indicating the loss of a classical near-wall turbulence structure. The absence of Reynolds shear stress has noticeably been observed in the experiments of Warholic \& Hanratty\cite{warholic1999idr} for their MDR flow. Even at a slightly lower drag reduction, the polymer shear stress dominates the Reynolds shear stress, which is consistent with HDR flows\cite{dubief2004cdr}.

The statistical analysis of our polymer flow simulations shows turbulent states predominantly supported by polymer stresses. The similarity between the two cases $Re=1000$, $Wi=100$ and  $Re=6000$, $Wi=700$ indicates that the state, EIT, observed at the lowest Reynolds numbers is likely to exist at higher Reynolds numbers, which suggests that EIT could be the asymptotic structural state of wall-bounded polymer flows.

\section{Topology of EIT}

\begin{figure}
\centerline{
\includegraphics[width=0.39\textwidth]{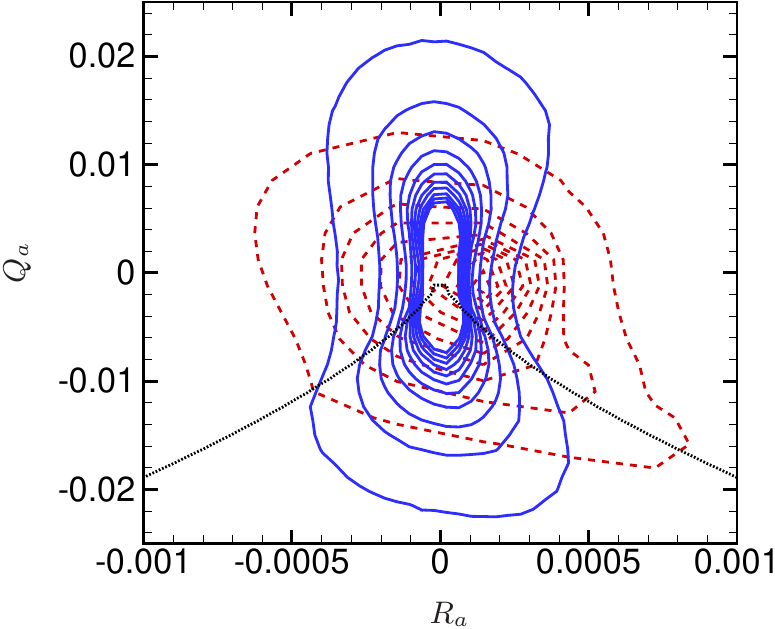}
}
\caption{\label{fig:topo} Joint probability density functions of flow topology in the $(Q_a,R_a)$ phase plot for polymeric flows at $Re=1000$ ({\color{blue}\solid}) and 6000 ({\color{red}\dashed}) for $Wi=100$ . Line \dotted describes $D=27/4R_a^2+Q_a^3=0$, the vertical bounds of the four quadrants of flow topology. 
}
\end{figure}

%%BW\begin{figure}
%%BW\vspace*{12pt}
%%BW\centerline{
%%BW\includegraphics[width=0.98\textwidth]{Fig2-bw.pdf}
%%BW}
%%BW\caption{\label{fig:topo} (a) Shows the joint probability density functions of flow topology in the $(Q_a,R_a)$ phase plot for polymeric flows at $Re=1000$ ({\solid}) and 6000 for $Wi=100$ ({\dashed}). Line \dotted describes $D=27/4R_a^2+Q_a^3=0$, the vertical bounds of the four quadrants of flow topology. Instantaneous contours of polymer extension in an $x-y$ plane showing streamwise sheet-like regions for $Re=1000$ (b) and $Re=6000$ (c); the continuous and dashed lines superimposed to the contours of polymer stretch represent isosurfaces of $Q_a$. Instantaneous isosurfaces of the second invariant of the velocity gradient tensor for $Q_a=\pm0.025$ at $Re=1000$ (d) and $Q_a=\pm0.25$ at $Re=6000$ (e). Gray: positive $Q_a$; black: negative $Q_a$. (Color online.)
%%BW}
%%BW\end{figure}

The flow topology is a critical component of the dynamics of polymers \citep{terrapon2004sps}, since it governs the stretching terms in the transport equation  of the conformation tensor, \eqnref{eq:C}, through the velocity gradient tensor $\vnabla \vu=\partial_ju_i$. We therefore apply the classical reduction of the flow into a joint probability density function (jpdf) of the second $Q_a=-\partial_ju_i\partial_iu_j/2$ and third $R_a=-\partial_ju_i\partial_ku_j\partial_iu_k/3$ invariants of $\vnabla \vu$\cite{soria1994study,ooi1999study}. Based on the sign of the discriminant $D=27/4R_a^2+Q_a^3$, quadrants I $(R_a>0,D>0)$ and II $(R_a<0,D>0)$ of \figref{fig:topo} define spiraling flows under compression and extension, respectively, and III $(R_a<0,D<0)$ and IV $(R_a>0,D<0)$ biaxial compressional and extensional flows, respectively. At $Re=6000$, the jpdf contours exhibit an inverted teardrop shape common to many turbulent flows, in particular of Newtonian turbulent channel flow. The topology distribution for the lower Reynolds number flow is significantly different with a quasi-symmetry around both $R_a=0$ and $Q_a=0$ and the confirmation of the existence of biaxial extensional events ($D<0$). 

\begin{figure}
\centerline{
\includegraphics[width=0.98\textwidth]{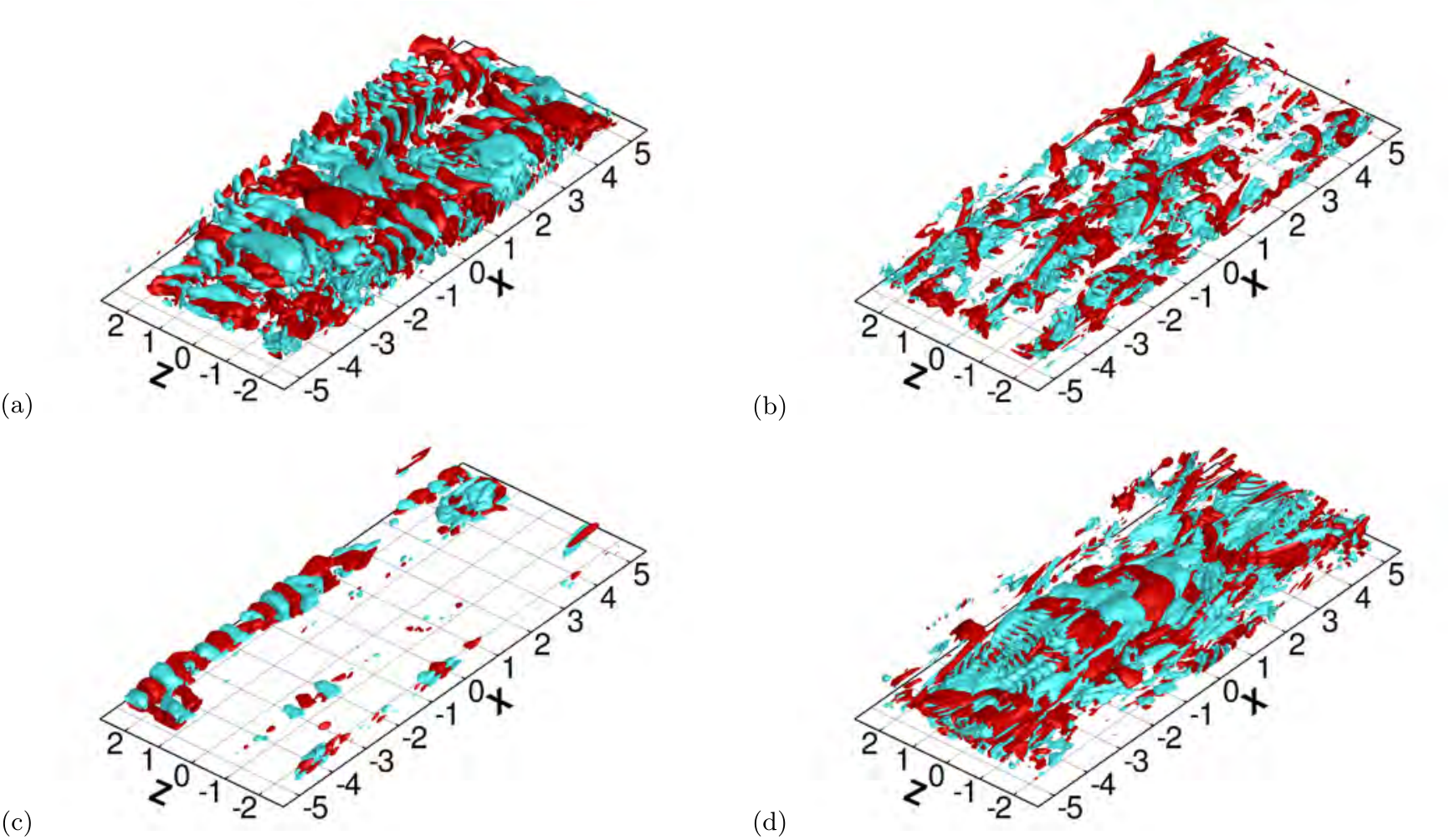}
}
\caption{\label{fig:Q-iso-all} Instantaneous isosurfaces of the second invariant of the velocity gradient tensor for $Q_a=\pm0.025$ at $Re=1000$, $Wi=100$ (a); $Q_a=\pm0.25$ at $Re=6000$, $Wi=100$ (b); $Q_a=\pm0.25$ at $Re=6000$, $Wi=700$ during hibernating (c) and active (d) state. Red: positive $Q_a$; cyan: negative $Q_a$. 
}
\end{figure}

Figure \ref{fig:Q-iso-all} shows the 3D structure of $Q_a$ with positive and negative isosurfaces of $Q_a$ for $Re=1000$ and 6000. At the lower Reynolds number (\figref{fig:Q-iso-all}a), the structure of the $Q_a$ field is predominantly spanwise with trains of cylindrical structures of various scales. The smallest scale structures depicted by the threshold chosen in this figure are organized in trains of structures of alternating sign.  Much larger scales are observed throughout the domain, with a typical spanwise dimensions comparable to the small scale structures. The typical streamwise dimension appears to be a couple of wavelength of the structures forming the trains of small scales. At $Re=6000$ and moderate elasticity ($Wi=100$), we chose the particular instant that is shown in \figref{fig:Q-iso-all}(b) because the structure is characteristic of an active state\cite{xi2010active}, or state of high drag (thus higher turbulence). No trains of alternating sign cylindrical structures of $Q_a$ may be observed. However, these trains reappear during hibernating states\cite{xi2010active} (not shown for $Wi=100$), as evident from \figref{fig:Q-iso-all}(c) for $Re=6000$ and $Wi=700$. Active states at $Wi=700$ (\figref{fig:Q-iso-all}d) consist also of hairpin vortices, though much larger than for $Wi=100$, and trains of small scale cylindrical structures of alternating sign of $Q_a$. Some of these trains appear in the wake of heads of hairpin vortices and were also observed at higher Reynolds numbers \citep{dubief2010polymer}. Other train occurrences are also observed in regions of low turbulence, consistent with the hibernating state's picture.

\begin{figure}
\vspace*{12pt}
\centerline{
\includegraphics[width=0.69\textwidth]{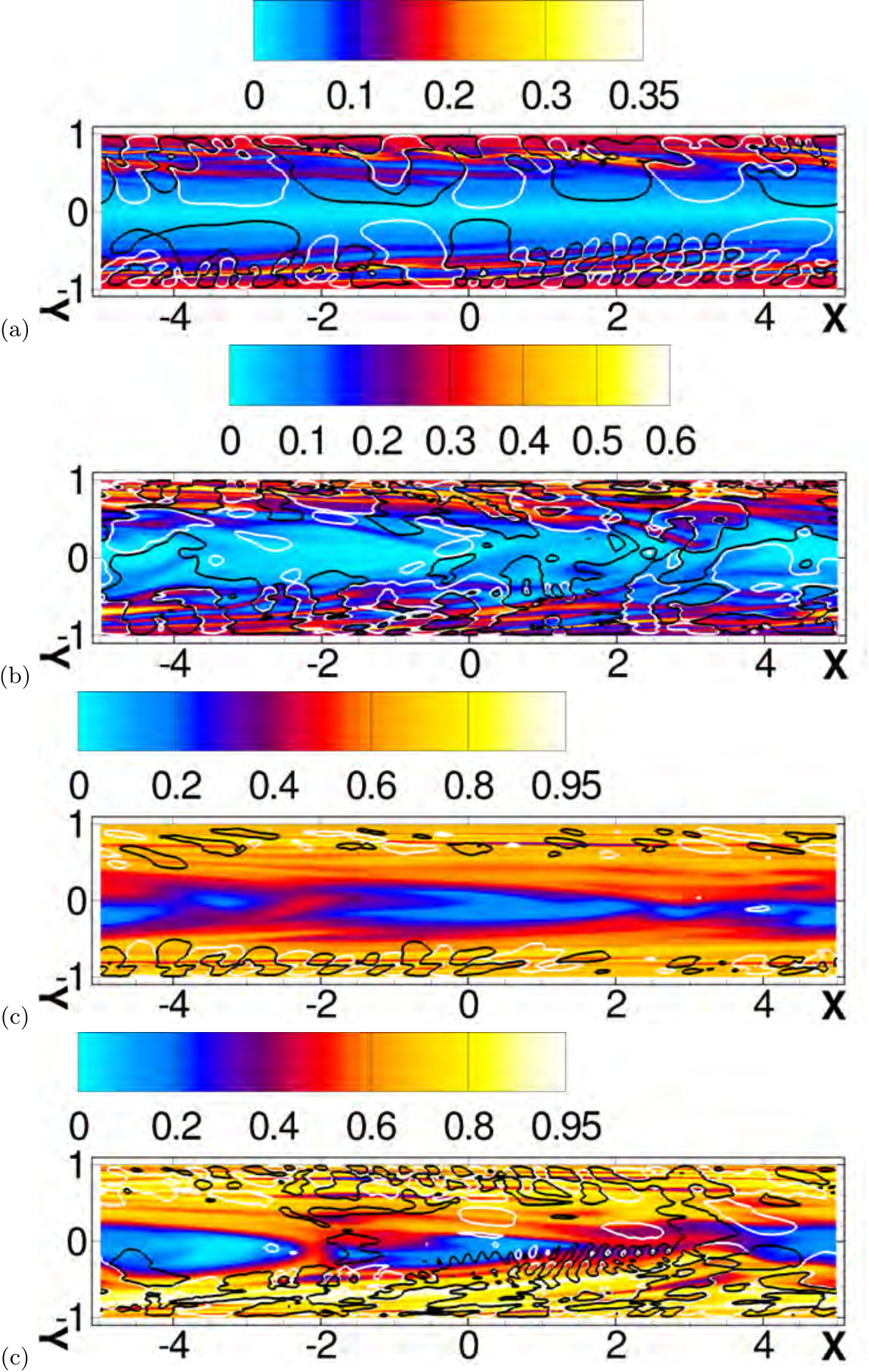}
}
\caption{\label{fig:polyext}  Instantaneous contours of polymer extension in an $x-y$ plane showing streamwise sheet-like regions. $Re=1000$, $Wi=100$ (a);  $Re=6000$, $Wi=100$ (b) and $Re=6000$, $Wi=700$ for hibernating (c) and active (d) state. The continuous and dashed lines superimposed to the contours of polymer stretch represent isosurfaces of $Q_a$: black, negative; white, positive.
}
\end{figure}

Local extensional flows are captured in contours of polymer stretch (\figref{fig:polyext}) in the form of thin sheets of locally high polymer stretch, tilted upwards and elongated in the flow direction. Superimposed to the contours of polymer stretch are the contours of positive and negative $Q_a$, which reveal the existence of trains of circular regions of alternating $Q_a$-sign associated to sheets of large polymer stretch. The intermittence of these sheets  is clear for $Wi=100$ at both Reynolds numbers, where the sheets of highly stretched polymers are surrounded by large regions of low stretch (Figures \ref{fig:polyext}a-b). The smallest scales (cylindrical structures of positive and negative $Q_a$) cluster along the sheets of highly stretched polymers. The cross-section of these cylindrical structures are shaped like the outside contour of a dumbbell, or an ellipsoid squeezed along its shorter radius. The large scales of the $Q_a$ field appear also to alternate sign along the streamwise direction. These large scale structures seem to arise from the smallest scales maybe through a merger of the upper head (or most outward head) of the small scale dumbbell structures. This is only a speculation at this point which warrants further investigation but is not critical to the current study. 

 In the active state of the $Re=6000$, $Wi=100$ simulation (\figref{fig:polyext}b), the trains of small scales are not present, at least in the cross-section considered here.  At the highest Weissenberg number, highly stretched polymers occupy most of the domain, yet sheets are still present in the hibernating and active states, Figures \ref{fig:polyext}(c) and (d), respectively. During the hibernating state, the structure of $Q_a$ exhibits the same dumbbell shape found at $Re=1000$.  The plane in Figure \ref{fig:polyext}(d) cuts through a hairpin vortex, illustrating one of the train of small-scale $Q_a$ structures spanning over $-1\lesssim x/h\lesssim 3$ in the lower channel half. 
 
One important conclusion of this topological study is that EIT consists of cylindrical structures of $Q_a$ of alternating signs at least at two scales, small dumbbell-like cylinders and larger structures that seem to results from the merging of the dumbbell cylinders. Another important observation is that the features of EIT disappear when the flow is too turbulent or the polymer solution not elastic enough. EIT reappears during hibernation events or even during active events if the Weissenberg number is sufficiently large. This suggests that EIT is an asymptotic state that should occur when the elasticity of the solution can efficiently control and contain the growth of turbulence.

\section{Similarity between EIT and 2D turbulence}

\subsection{Energy transfer between polymers and turbulence}

\begin{figure}
\vspace*{12pt}
\centerline{
\includegraphics[width=0.49\textwidth]{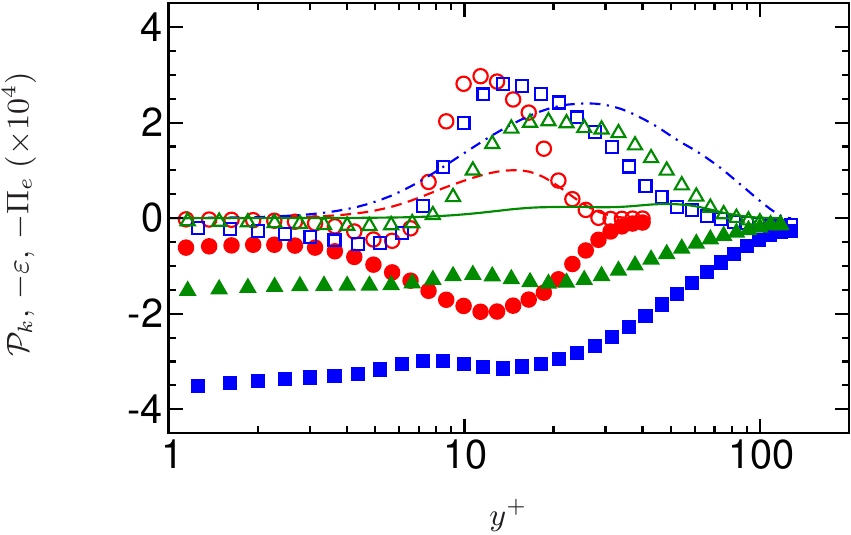}
}
\caption{\label{fig:energy} Profiles of the production of turbulent kinetic energy, ${\cal P}_k$ (lines), the negative contribution of dissipation rate of TKE, $\varepsilon$(closed symbols) and transfer of energy between elastic energy and TKE, $\Pi_e$ (open symbols). The sum of the volumetric integral of these term results in the energy balance found in \eqnref{eq:energy}. $Re=1000$, $Wi=100$: {\color{red} \dashed, \circle, \solidcircle}; $Re=6000$, $Wi=100$: {\color{blue} \chndot, \square, \solidsquare}; $Re=6000$, $Wi=700$: {\color{ForestGreen} \solid, \trian, $\solidtrian$}. }
\end{figure}

The only evidence presented so far for the sustainability of EIT through energy transfer from polymers to turbulence is the very existence of EIT at a Reynolds number much lower than the critical Reynolds number for Newtonian flows.  Following Dallas \etal\cite{dallas2010strong}, we use the integral of the budget of turbulent kinetic energy (TKE):
\begin{equation}
\int_V{\cal P}_k\,dV-\int_V\varepsilon\,dV-\int_V\Pi_e\,dV=0,\label{eq:energy}
\end{equation}
where $V$ is the volume of the channel. The three terms involved are the production of TKE, the dissipation rate of TKE and the energy transfer between polymers and TKE, which are respectively defined as:
\begin{subequations}
\label{eq:energyterms}
\begin{eqnarray}
{\cal P}_k&=&-\avg{u'v'}\der{U}{y}\label{eq:prod}\\
\varepsilon&=&\frac{2\beta}{Re}\avg{\tS':\tS'}\label{eq:eps}\\
\Pi_e&=&\frac{1-\beta}{Re}\avg{\tS':\tT'}\label{eq:Pie}
\end{eqnarray}
\end{subequations}
The tensor $\tS'$ is the fluctuating strain rate tensor $\tS'=(\vnabla\vu'+\vnabla\vu'^T)/2$ and $\tT'$ is the fluctuating polymer stress tensor.
\figref{fig:energy} shows the profile across one half of the channel of the three terms of interest to for polymer flows at $Re=1000$ and 6000. The wall-normal coordinate is normalized by viscous scales. In \eqnref{eq:energy}, the production and dissipation rate of TKE terms are always positive. The energy transfer between polymer and turbulence divides the flow in two regions. In a thin region attached to the wall ($y^+\lesssim 7$, polymers supplements the dissipation rate of TKE. Beyond this ``polymeric viscous sublayer'', polymers support and even create, for $Re=1000$, turbulence with transfers from elastic to turbulent kinetic energy. Lending further credence to our proposition that EIT is the asymptotic state of polymer drag reduction, the production of TKE in the highest Weissenberg and Reynolds number flow is overwhelmingly dominated by $-\Pi_e$, in a similar fashion as for the $Re=1000$ flow.

\figref{fig:energy} brings further support to de Gennes's theory\cite{de1990introduction} of exchange of energy between elastic energy and TKE. A similar point was made by Dallas \etal\cite{dallas2010strong} using DNS as well. The picture of EIT that emerges so far is one of extraction of TKE through the damping mechanism of coherent structures\cite{dubief2004cdr} and injection of elastic energy into TKE through elastic instabilities caused by the formation of thin sheets of polymer stretch, as highlighted in the previous section and \figref{fig:polyext}.

\subsection{The 2D turbulence analogy}

\begin{figure}
\centerline{
\includegraphics[width=0.98\textwidth]{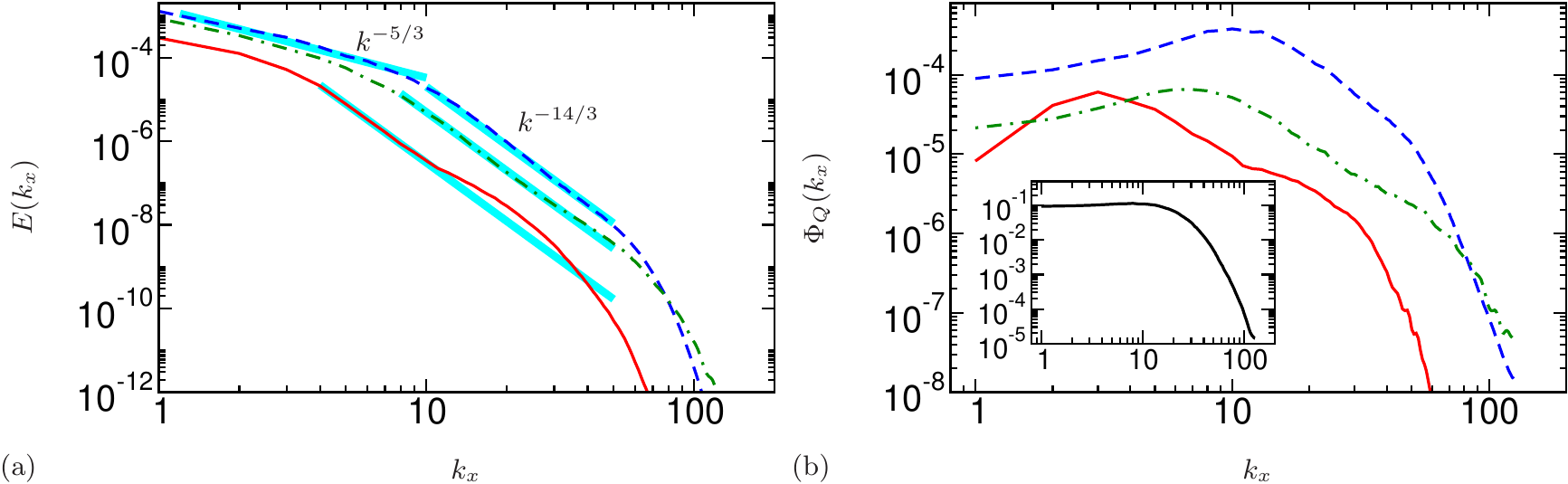}
}
\caption{\label{fig:spec} (a) Spectra of turbulent kinetic energy at $y^+=15$. (b) Spectra of $Q_a$ at the same location. {\color{red} \solid}: $Re=1000$, $Wi=100$; {\color{blue} \dashed} $Re=6000$, $Wi=100$; {\color{ForestGreen} \chndot} $Re=6000$, $Wi=700$. Inset: Spectrum of $Q_a$ for Newtonian turbulence at $Re=6000$, same location.
}
\end{figure}

Equipped with the knowledge of energy transfer from polymers to TKE, we now turn to spectral analysis for a discussion on the mechanism of such transfer. An interesting characteristic of the spectral distribution of TKE in EIT flow is the observation of $k^\alpha$ behavior at high wavenumbers. \figref{fig:spec}(a) shows longitudinal 1D spectra of TKE at $y^+=15$ for all polymeric simulations. The location was chosen for its good approximation of the maximum energy transfer of elastic energy to TKE for all simulations. In spite of the low Reynolds numbers considered here, a range of wavenumbers, between $k_x\approx4$ to 10 for $Re=1000$ and $k_x\approx 8$ to 50 for $Re=6000$, show a power decay that fits well $k^{-14/3}$. We checked the robustness of the power law behavior by plotting the exponent derived from $\alpha(k_x)=(k_x/E(k_x))(d E/dk_x)$ and confirming the existence of a plateau of $\alpha(k_x)$ in the range of wavenumbers of interest (not shown). The $-14/3$ exponent bears, at this point, no theoretical significance; the best fit was found to be around 4.6. In the two EIT flows, the higher wavenumbers depart  from $-14/3$ and seem closer to $-11/3$ (not shown). Nonetheless the important observation here is that the power law decay is between $-3$ and $-5$.  \figref{fig:spec}(b) correlates the lower bound wavenumber of the $-14/3$ region with the maximum spectral contribution to the fluctuations of $Q_a$. This energetic peak is dramatically more marked in polymeric flows than in Newtonian turbulence, and corresponds to the trains of cylindrical structures observed in the instantaneous visualizations shown in Figures \ref{fig:Q-iso-all} and \ref{fig:polyext}. For reference, $k^{-5/3}$ is also fitted to the lowest wavenumbers. The reasonable visible agreement is more an effect of the vertical scale of the plot rather than an actual $-5/3$ behavior, yet this reference will help the present discussion.

\begin{figure}
\centerline{
\includegraphics[width=0.98\textwidth]{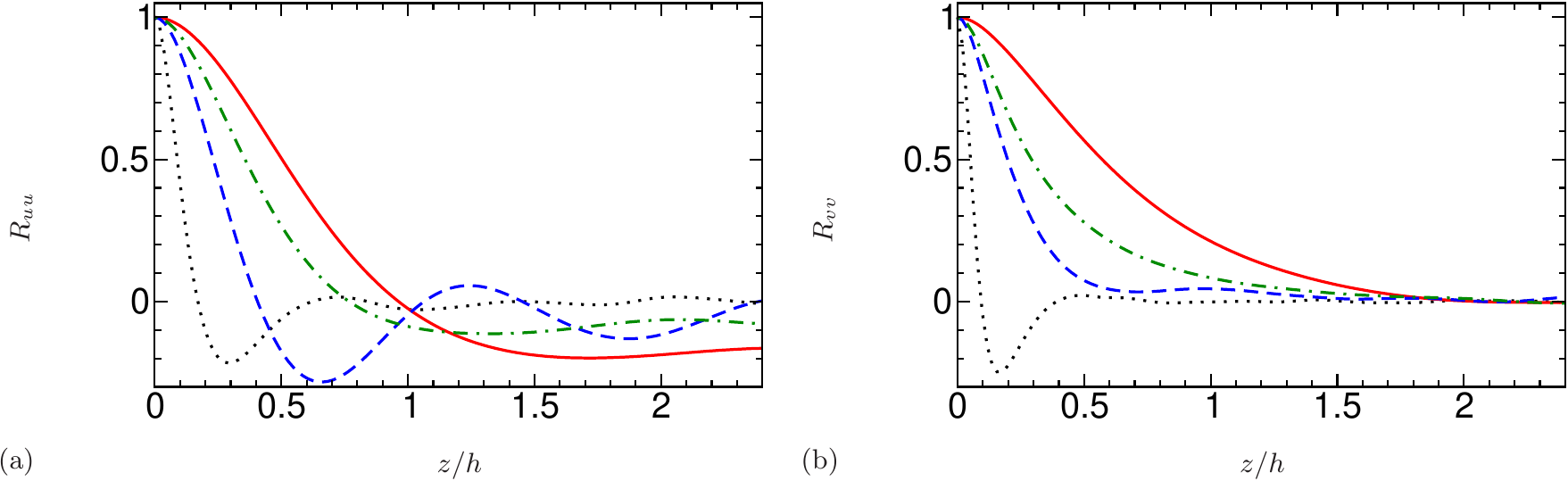}
}
\caption{\label{fig:corr} Spatial correlations in the spanwise direction of streamwise (a) and wall-normal (b) velocity fluctuations at $y^+=15$. Line captions for polymeric flow is the same as in \figref{fig:spec}. \dotted, $Re=6000$ Newtonian;  
}
\end{figure}

The spectral behavior described here provides circumstantial evidence of a 2D-like dynamics of polymeric flows at or close to EIT. Indeed, 2D flows are known for their clearly observed, documented and even theoretically described inverse energy cascade\cite{boffetta2012two}. In fact, the best analogy between our simulations and 2D turbulence is found in numerical experiments where friction is introduced\cite{nam2000lagrangian}. In such experiments, the flow is governed by the following incompressible, 2D Navier Stokes equation:
\begin{equation}
¥\pdert{\vu}+(\vu\cdot\vnabla)\vu=-\vnabla p+\nu\nabla^2\vu-\nu_f\vu+\vex{f}\,,\label{eq:2dmom}
\end{equation}¥
where $\nu_f$ is the friction coefficient and $\vex{f}$ is a discrete random forcing at a given intermediate wavenumber $k_d$. These flows are solved in periodic domains.  
Such flows experiences an inverse energy cascade as well\cite{nam2000lagrangian}, where energy cascades toward the largest scales ($k<k_d$) with an energy spectrum of the form:
\begin{equation}
E(k)=C_K\varepsilon^\frac{2}{3}k^{-\frac{5}{3}} \text{ for }k<k_d\,.
\end{equation}
At larger wavenumbers, energy is dissipated through an enstrophy cascade, where the enstrophy is the integral of the square of the vorticity. The influence of the friction coefficient is felt in the exponent of the spectral decay of enstrophy cascade which corresponds to
\begin{equation}
E(k)\propto k^{-3-\xi}\,,
\end{equation}
where $0\leq\xi\leq2$, in agreement for the slopes found in our polymeric simulations. The analogy is further supported by the significant increase of the correlation length in the spanwise direction as shown in \figref{fig:corr} for the streamwise and wall-normal velocity fluctuations. For the lowest Reynolds number simulation and the highest elasticity at $Re=6000$, the correlation length is dramatically increased. The flow does not experience however a reduction in dimensionality as dramatic as the ones observed in strongly rotational flows\cite{hossain1994reduction} for instance. It is however striking to note that, even at the lowest Reynolds number, the structure depicted by the correlation of the streamwise velocity fluctuations indicate the existence of streak-like structures, yet much wider than for the Newtonian flow. This was confirmed by instantaneous visualizations of wall shear stress (not shown).

The concept of inverse energy cascade does however makes sense in the context of our polymeric flows. There is an obvious resemblance between the polymeric flow momentum equation \eqnref{eq:mom} and the 2D equation \eqnref{eq:2dmom} in the dynamics at play, where the divergence of polymer stress embodies both the friction term and the injection of energy term. Since the same term provides large scale dissipation (drag reduction mechanism) and energy injection, one may expect differences between the two dynamical systems, which could explain the coexistence of the -14/3 and -11/3 in EIT flows. 

The elastic instabilities captured through isosurfaces and contours of $Q_a$ (Figures \ref{fig:Q-iso-all}, \ref{fig:polyext} and \ref{fig:spec}b) occur at  intermediate  wavenumbers close to the lower bound of the -14/3 range shown in \figref{fig:spec}.  Whether such instabilities excite or directly create larger scales is a question we cannot answer with the present analysis. One possible approach that could provide valuable insights in the future is the method proposed by Biferale \etal\cite{biferale2012inverse} which demonstrated the existence of inverse energy cascade in 3D isotropic turbulence. Incidentally, Biferale {\etal} demonstrated that ``2D and 3D properties naturally coexist in all flows in nature''. The interesting conjecture derived from their study and our observations here is that a flow that is locally two-dimensional may exhibit a behavior akin to that observed in true 2D turbulence, in particular an inverse energy cascade.  

\subsection{An instantaneous picture of transfer between elastic and turbulent kinetic energies}

\begin{figure}
\centerline{
\includegraphics[width=0.59\textwidth]{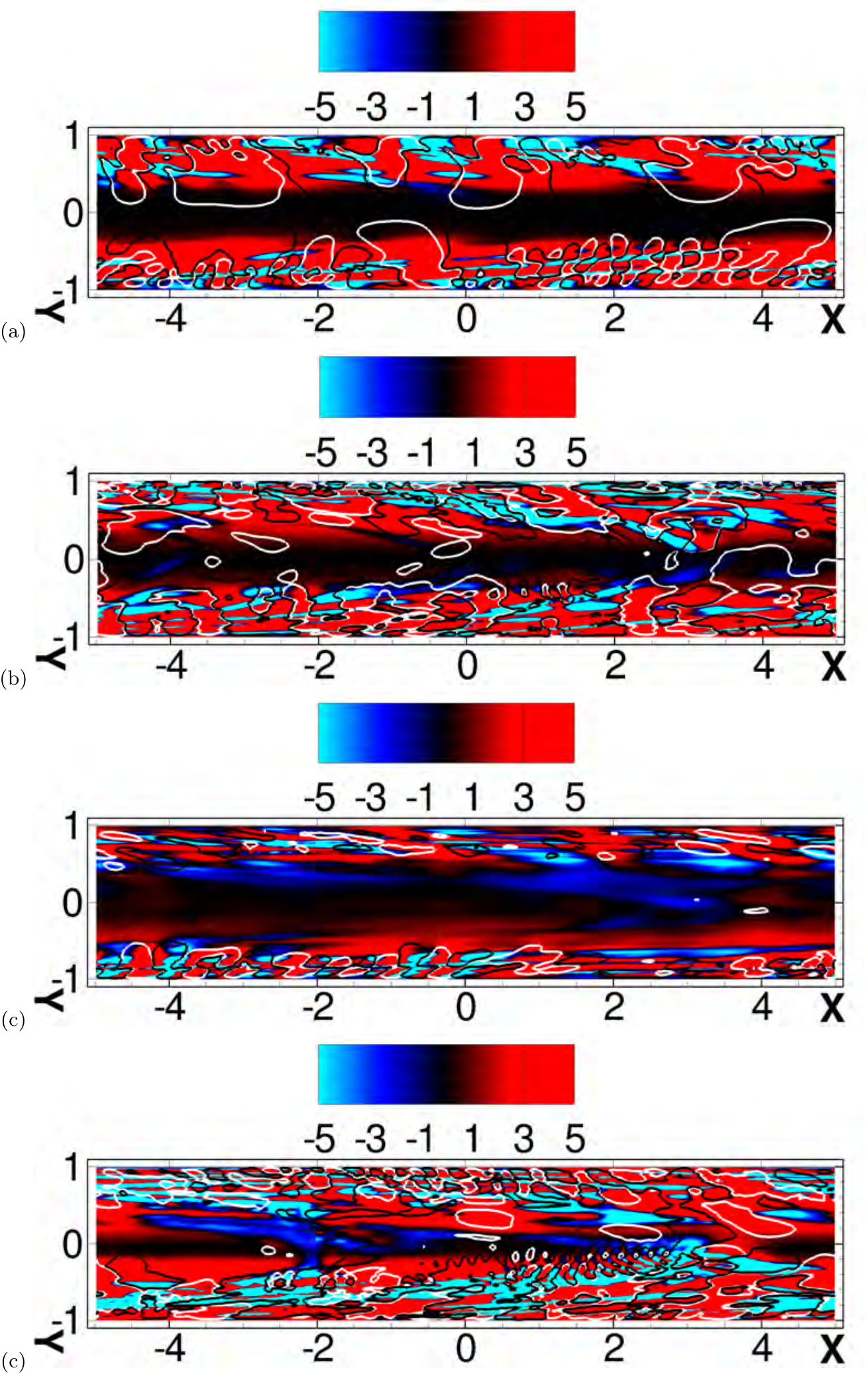}
}
\caption{\label{fig:Pie} Spatial correlations of streamwise (a) and wall-normal (b) velocity fluctuations in the spanwise direction. \solid, $Re=6000$ Newtonian;  
}
\end{figure}
 Assuming that the proposed analogy between EIT and 2D turbulence holds, the energy transfer term between polymers and TKE can be interpreted as follows. 
TKE is a large scale quantity, whereas the organization of polymers in sheets strongly suggests that polymers are intermediate or small scales, as shown by the $Q_a$ structures they create. Indeed the thickness of the sheets of highly stretched polymers appear to govern EIT, as evident from the clustering of $Q_a$ structures on these sheets.  Defining the volumetric integral as $\spavg{\bullet}=\int_V\bullet dV$,  the rate of change of TKE is thus
\begin{equation}
\der{\spavg{e_k}}{t}=\spavg{{\cal P}_k}-\spavg{\varepsilon}-\spavg{\Pi_e}
\end{equation}
and $-\spavg{\varepsilon}$ is always negative thus a sink of TKE. The scalar product between tensors $\tT'$ and $\tS'$ is effectively the rate of work done by the strain rate on the polymer stress. When fluctuations of polymer and strain tensors align, $\tT':\tS'>0$, the work is a sink of turbulent kinetic energy, consistent with the increase of polymer stress in sheets (extensional flows) which triggers large increases in extensional viscosity. Following our analogy, this process would describe a forward energy cascade. When polymer stress fluctuations oppose flow deformation fluctuations, $-\Pi_e$ becomes positive and contributes directly to the TKE, which amounts to a backward energy cascade. The interplay between $-\Pi_e$ and coherent structures isolated by isosurfaces of $Q_a$ is shown in \figref{fig:Pie}. The region of forward cascades are sheets, that one can visually correlate to sheets of highly-stretched polymers from \figref{fig:polyext}. These regions also appear to be the cause of the narrow part of the dumbbell shape of the $Q_a$-isosurfaces (\figref{fig:Pie}a for $0\lesssim x/h\lesssim4$, for instance). The backward energy cascade surrounds the sheets of forward cascade and contain the largest scales of $Q_a$. Note that in the active state depicted in \figref{fig:Pie}(d), one can observe the wake of the head of a hairpin vortex where the flow is extensional and triggers EIT. The distribution of backward and forward energy cascade is evident yet very complex and warrants future studies. Even at the lowest drag reduction studied here, there is still significant amount of backward energy cascade.

%One may speculate that the physical mechanisms of inverse energy transfer of elastic energy to TKE have some relation with the mechanisms of 2D turbulence inverse energy cascade based on the circumstantial evidences presented so far. The justification for this analogy is the quasi two-dimensionality of the flow, observed in instantaneous visualizations of flow structures and spatial correlations. Indeed, when the flow is more turbulent ($Re=6000$, $Wi=100$), the spectra show less obvious enstrophy-cascade type behavior and less evidence of trains of $Q_a$ cylindrical structures, the hallmark of EIT. EIT seems thus to need and thrive on a partial and local reduction of dimensionality. 

%Now we direct the attention of the reader to the range of low wavenumbers with a $k^{-1}$ behavior. This range is much larger for the Newtonian flow than for any of the polymeric flows. In the log-region of the Newtonian flow, the $k^{-1}$ was related to the logarithmic behavior of the mean velocity profile by Perry \& Chong\cite{perry1982mechanism} from their attached-eddy hypothesis. The absence of log behavior reported by White \etal\cite{white2012re} can be now correlated to a significant shrinking of the range of wavenumbers approaching a $k^{-1}$ behavior.

\section{Proposed mechanism of EIT}

\begin{figure}
\centerline{
\includegraphics[width=0.69\textwidth]{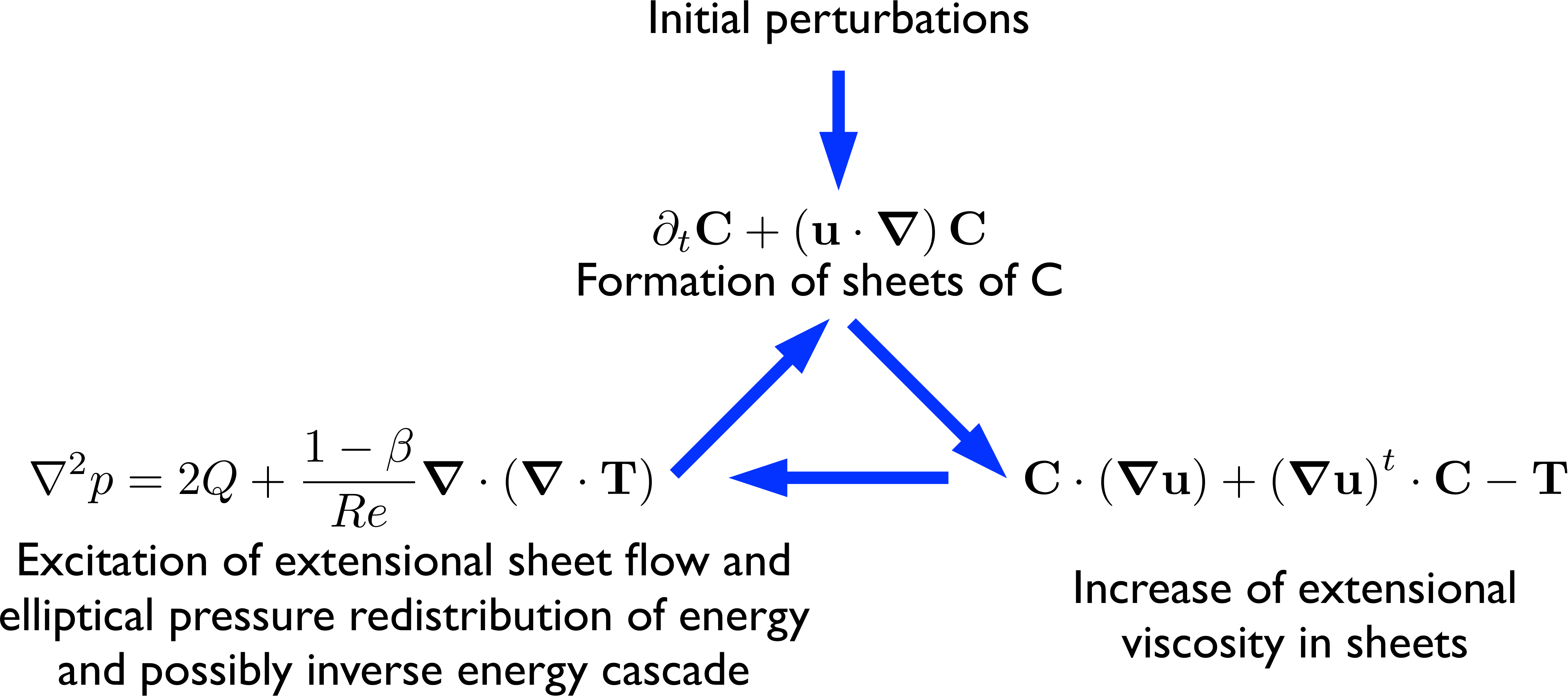}
}
\caption{\label{fig:sketch} Conceptual skecth of the mechanism of EIT. 
}
\end{figure}
The dynamic of EIT was found to be consistent across the two regimes shown in \figref{fig:stat}(a) ($Re<Re_c$ and $Re>Re_c$), and can be best described by (i) taking the divergence of \eqnref{eq:mom}, which yields the Poisson equation for pressure in a viscoelastic flow:
\begin{equation}
\nabla^2p=2Q_a-\frac{1-\beta}{Re}\vnabla\cdot(\vnabla\cdot\tT), \label{eq:D2p}
\end{equation}
and (ii) considering the hyperbolic nature of the transport equation of the conformation tensor, \eqnref{eq:C}, caused by the absence of diffusion \citep{dubief2005nai}. For $Re<Re_c$, a perfectly laminar flow stretches polymers through the action of shear (stretching term in Eq. \ref{eq:C}). The introduction of small perturbations into the flow excites the unstable nature of the nonlinear advection term $(\vu\cdot\vnabla)\tC$, resulting in the formation of sheets or cliffs of polymer stretch akin to cliffs of scalar concentration observed in the turbulent transport of low-diffusivity passive scalar \citep{schumacher2005very}. This behavior is obvious when comparing \figref{fig:topo}(b-c) with LIF (Laser Induced Fluorescein) images of fluorescein dye concentration in polymer drag-reduced wall bounded flows \citep{somandepalli2010concentration}.  The sheets of high polymer stretch, hosting a significant increase in extensional viscosity, create a strong local anisotropy, with a formation of local low-speed jet-like flow.  The response of the flow is through pressure (Eq. \ref{eq:D2p}), whose role is to redistribute energy across components of momentum, resulting in the formation of waves, or trains of alternating rotational and straining motions as shown by the $Q_a$ isosurfaces. The mechanism shares some similarity with the Kelvin-Helmholtz instability, except that the thickness of these sheets is too close to the Kolmogorov scale (smallest scale of turbulence) for vortices to be created. Once triggered, EIT is self-sustained since the elastic instability creates the very velocity fluctuations it feeds upon. Interestingly, Samanta \etal\cite{samanta2012transition} and Dubief \etal\cite{dubief2010polymer} show that the phenomenon of EIT is not confined to low Reynolds numbers, unlike elastic turbulence \citep{groisman2000elastic}.
We also presented circumstantial evidences that EIT shares some interesting similarities with 2D turbulence, which could provide a reasonable framework for further studies. The mechanism of EIT is summarized in \eqnref{fig:sketch}.

\section{Conclusion}

Elasto-inertial turbulence offers a new perspective on polymer drag reduction. First, it provides support to DeGennes' \cite{de1990introduction} theory of energy transfers between polymers and flow. Second, EIT allows us to consider the possible structure of MDR for very large elasticity ($Wi\rightarrow \infty$): The sheer magnitude of extensional viscosity is likely to prevent the emergence of any vortical structures, thus leaving MDR to be sustained by near-wall spanwise structures similar to the ones observed at low Reynolds numbers (\figref{fig:topo}d). The absence of vortices, however, does not signify the absence of streaks. Indeed, there is no apparent means in our current understanding of EIT to damp streaks. Our study suggests that, given the right amount of injected energy, the streak instability can be triggered and sustained at subcritical Reynolds numbers.

The asymptotic state of polymer drag reduction in the absence, or near absence, of vortices should therefore be driven by the nonlinear transport of polymer stretch, which resembles the transport of a high Schmidt number passive scalar, and the response of the flow to a sheet-like, strongly anisotropic field of effective viscosity governed by the extensional viscosity of polymers. As discussed by Dubief \etal\cite{dubief2010polymer,dubief2012polymer}, the flow is therefore stuck in a transitional state, specifically the stage of breakdown of nonlinear flow instabilities, which does not support a logarithmic mean velocity profile \citep{klewicki2011mean}. Unfortunately, a major obstacle in the derivation of a low Reynolds (or possibly high Reynolds) number correction of Virk's log law is the lack of   theoretical understanding of high Schmidt number active scalar transport in nonlinear anisotropic flows. Our analogy with 2D turbulence speculates that the mechanism by which turbulence is supported may be that of an inverse energy cascade. The potential broader impact of this discussion could be that inverse energy cascade in 3D turbulent flows may be related to the local reduction dimensionality of the flow. Another perspective is also the possibility to develop a theoretical approach using a 2D or quasi 2D flows.

Future research should therefore focus on the dynamics of active scalars and the backscatter of energy caused by  polymer dynamics. Another interesting future research direction is whether the action of the polymers in the flow is catalytic or one of direct energy exchange between polymer and flow instabilities. At low Reynolds numbers, the process appears to be a direct energy exchange; however, certain aspects of EIT at high Reynolds numbers may be catalytic, e.g. the trigger of elastic instability through bypass transition.

\begin{acknowledgments}
The Vermont Advanced Computing Center is gratefully acknowledged for providing the computing resources necessary for our simulations. YD acknowledges the support of grant No. P01HL46703 (project 1) from the National Institutes of Health. VET acknowledges the financial support of a Marie Curie FP7 Career Integration Grant within the 7th European Community Framework Programme (Grant Agreement n$^\circ$ PCIG10-GA-2011-304073). JS acknowledges the support of the Australian Research Council.
\end{acknowledgments}

\bibliography{dubief-terrapon-soria}


% Produces the bibliography via BibTeX.

\end{document}